\documentclass[11pt]{article}

\usepackage{graphicx}
\usepackage{rotating}
\usepackage{amssymb}
\usepackage{mathptmx}
\usepackage[compact]{titlesec} 

\usepackage{comment}
\def\pname{{\sc{Skip}}}
\def\lekid{{\sc{Lekid}}}
\def\lekids{{\sc{Lekid}s}}

\begin{document}

\title{The Detector System for the Stratospheric Kinetic Inductance
  Polarimeter (\pname)}

\author{
B.~R.~Johnson$^{1}$,
P.~A.~R.~Ade$^{4}$,
D.~Araujo$^{1}$,
K.~J.~Bradford$^{1}$,
D.~Chapman$^{1}$, \\
P.~K.~Day$^{7}$,
J.~Didier$^{1}$, 
S.~Doyle$^{4}$,
H.~K.~Eriksen$^{5}$,
D.~Flanigan$^{1}$,
C.~Groppi$^{2}$, \\
S.~Hillbrand$^{1}$,
G.~Jones$^{6,1}$,
M.~Limon$^{1}$,
P.~Mauskopf$^{2}$,
H.~McCarrick$^{1}$, \\
A.~Miller$^{1}$,
T.~Mroczkowski$^{3}$,
B.~Reichborn-Kjennerud$^{1}$,
B.~Smiley$^{1}$, \\
J.~Sobrin$^{1}$,
I.~K.~Wehus$^{7}$,
J.~Zmuidzinas$^{3,7}$ \\ \\ 
\textit{1)~Department of Physics, Columbia University, New York, NY 10027, USA} \\
\textit{2)~Department of Physics, Arizona State University, Tempe, AZ 85287, USA} \\
\textit{3)~Department of Physics, Caltech, Pasadena, CA 91125, USA} \\
\textit{4)~School of Physics and Astronomy, Cardiff University, Cardiff, Wales, CF243YB, UK} \\
\textit{5)~Institute of Theoretical Astrophysics, University of Oslo, N-0315 Oslo, Norway} \\
\textit{6)~National Radio Astronomy Observatory, Charlottesville, VA 22903, USA} \\
\textit{7)~Jet Propulsion Laboratory, Caltech, Pasadena, CA 91109, USA} \\
}

\date{July 31st, 2013}

\maketitle


\begin{abstract}

The Stratospheric Kinetic Inductance Polarimeter (\pname) is a
proposed balloon-borne experiment designed to study the cosmic
microwave background, the cosmic infrared background and Galactic dust
emission by observing 1133 square degrees of sky in the Northern
Hemisphere with launches from Kiruna, Sweden.  The instrument contains
2317 single-polarization, horn-coupled, aluminum lumped-element
kinetic inductance detectors (\lekids).  The \lekids\ will be
maintained at 100~mK with an adiabatic demagnetization refrigerator.
The polarimeter operates in two configurations, one sensitive to a
spectral band centered on 150~GHz and the other sensitive to 260 and
350~GHz bands.  The detector readout system is based on the ROACH-1
board, and the detectors will be biased below 300~MHz.  The detector
array is fed by an F/2.4 crossed-Dragone telescope with a 500~mm
aperture yielding a 15~arcmin FWHM beam at 150 GHz.  To minimize
detector loading and maximize sensitivity, the entire optical system
will be cooled to 1~K.  Linearly polarized sky signals will be
modulated with a metal-mesh half-wave plate that is mounted at the
telescope aperture and rotated by a superconducting magnetic bearing.
The observation program consists of at least two, five-day flights
beginning with the 150~GHz observations.


\end{abstract}


\begin{figure}[ht]
\centering
\begin{minipage}[c]{0.55\textwidth}
\centering
\begin{small}
\begin{tabular}{|l|c|c|}
\hline
Instrument Configuration & Flight 1 & Flight 2 \\
\hline
Spectral Band Centers [GHz]                & 150        & 267, 350   \\
Spectral Bandwidth [$\delta\nu$/$\nu$]     & 0.27       & 0.22, 0.30 \\
Number of Detectors                        & 2317       & 1655, 662  \\
Total Number of Detectors                  & 2317       & 2317       \\
Detector NET [$\mu$K$\sqrt{\mbox{sec}}$]   & 58.4       & 149, 354   \\
Instrument NET [$\mu$K$\sqrt{\mbox{sec}}$] & 1.24       & 3.76, 14.1 \\
Aperture Diameter [mm]                     & 500        & 280        \\
F/\#                                       & 2.4        & 4.3        \\
Beam FWHM [arcmin]                         & 14.5       & 14.5, 11.3 \\
Total Sky Coverage [deg$^2$]               & 1133       & 1133       \\
$\ell$ Range                               & 10 to 1000 & 10 to 1000 \\
Flight Duration [days]                     & 5          & 5          \\
Time on CMB Patch [hours]                  & 102.7      & 102.7      \\
Duty Cycle [\%]                            & 86         & 86         \\
T sensitivity per 1$^{\circ}$ pixel [$\mu$K] & 0.069      & 0.21, 0.78 \\
Q sensitivity per 1$^{\circ}$ pixel [$\mu$K] & 0.097      & 0.29, 1.1 \\
\hline
Minimum~$r$~(99\% confidence)              & \multicolumn{2}{|c|}{0.02} \\
\hline
\end{tabular}
\end{small}
\end{minipage}
\begin{minipage}[c]{0.44\textwidth}
\centering
\includegraphics[width=\textwidth]{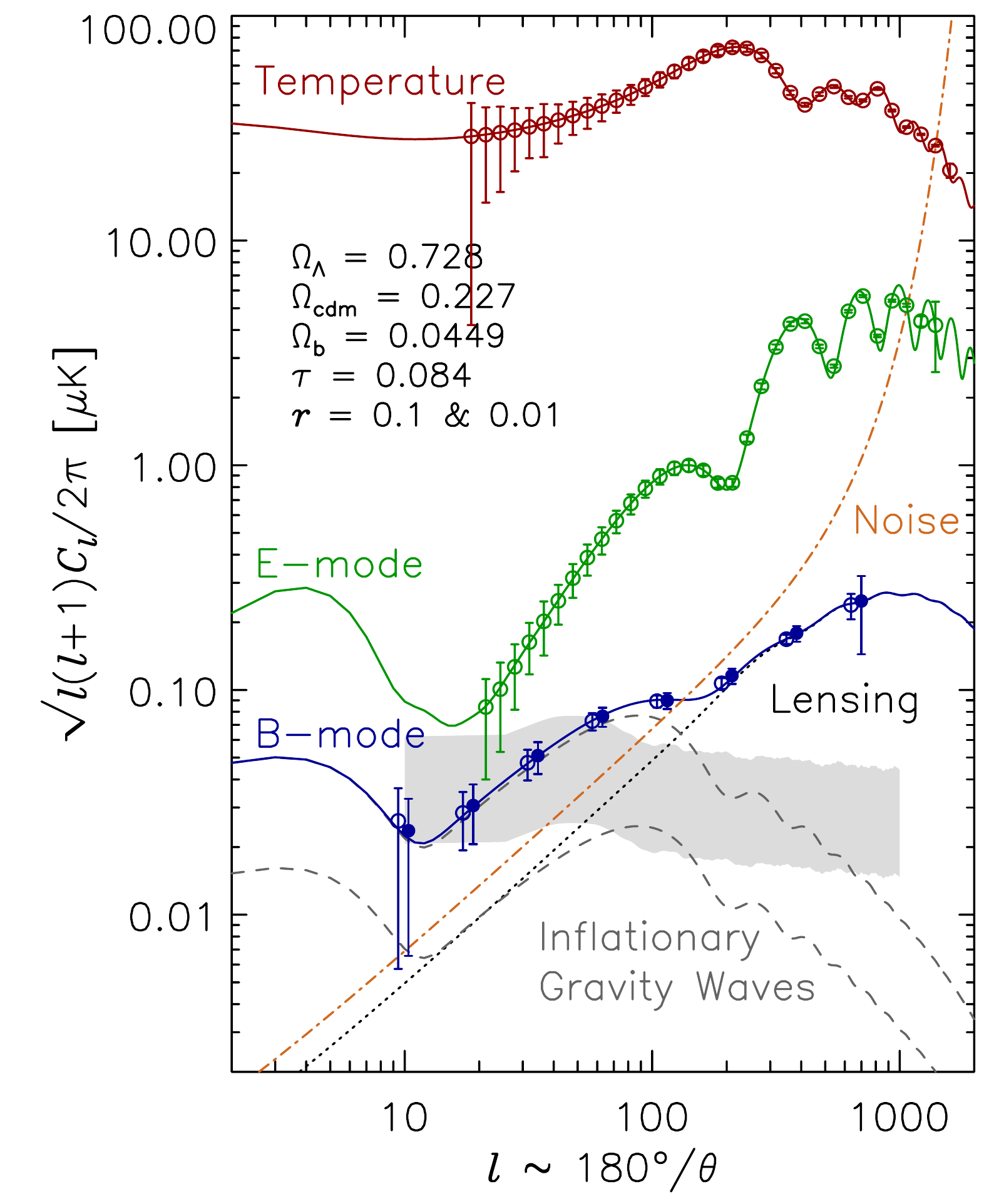}
\end{minipage}
\caption{\footnotesize{{\bf Left:} Key \pname\ performance
    characteristics. {\bf Right:} Theoretical angular power spectra
    for the CMB signals and the expected \pname\ performance. For the
    B-modes, the open circles show the raw sensitivity of the 150~GHz
    flight alone ignoring foreground component separation, while the
    closed circles include sensitivity degradation due to component
    separation.  The grey region shows the spectrum of foreground
    signals from Galactic dust assuming the polarization fraction is
    between $5\%$ and $15\%$ as predicted\cite{finkbeiner:1999}.}}
\label{fig:overview}
\end{figure}


\section{Introduction}

The cosmic microwave background radiation (CMB) carries an image of
the universe as it was 380,000 years after the Big Bang.
Measurements of the angular intensity and polarization anisotropies in
the CMB have proven to be a treasure trove of cosmological
information.
For example, CMB measurements have helped reveal that spacetime is
flat, the universe is 13.8 billion years old, and it is dominated by
cold dark matter and dark energy.
A faint and yet-to-be-detected divergence-free ``B-mode'' polarization
anisotropy signal in the CMB should contain additional information;
B-modes generated when CMB photons are gravitationally lensed by
large-scale structure will yield the values of physical parameters
such as the sum of the neutrino masses, and if discovered, an
inflationary gravity-wave (IGW) B-mode signal would reveal the energy
scale at which inflation occurred\cite{baumann:2009}.
\pname\ is a proposed balloon-borne experiment designed with the sky
coverage, angular resolution, sensitivity and spectral coverage needed
to either measure or constrain the amplitude of the IGW signal to
approximately the foreground confusion limit set by the aforementioned
gravitational lensing signal.
The table in Figure~\ref{fig:overview} shows the experiment
characteristics and the plot shows the forecasted CMB power spectra
results that will come from \pname.

To isolate the faint B-mode signals, brighter foreground signals, such
as Galactic dust emission, must be precisely measured and removed from
the CMB maps.
To disentangle the various sky signals, \pname\ will observe a
low-dust region in the Northern Hemisphere with three spectral bands,
and then fit a frequency-dependent parametric dust model to the data
pixel-by-pixel using standard likelihood
techniques\cite{eriksen:2008}.
The target sky patch is accessible with launches from Kiruna, Sweden.

Because the Galactic dust emission is so faint in the \pname\ patch,
the 350~GHz spectral band will also be able to image angular
fluctuations in the cosmic infrared background
(CIB)\cite{planck_cib:2013}.
The CIB was emitted by dust grains in distant galaxies, and it has a
redshifted, dust-like frequency spectrum.
The CIB probes the matter distribution on redshifts $z~\le~6$, rather
than at $z~\sim~1100$ like the CMB, so it can be used to constrain
star formation history and dark matter distribution.

Finally, \pname\ will investigate whether lumped-element kinetic
inductance detectors (\lekids) are a candidate detector technology for
a future CMB satellite mission.
The \pname\ instrument is described in Section~\ref{sec:instrument},
the detector system is described in Section~\ref{sec:detectors}, and
the observation program is described in Section~\ref{sec:observation}.


\begin{figure}[t]
\centering
\includegraphics[width=0.9\textwidth]{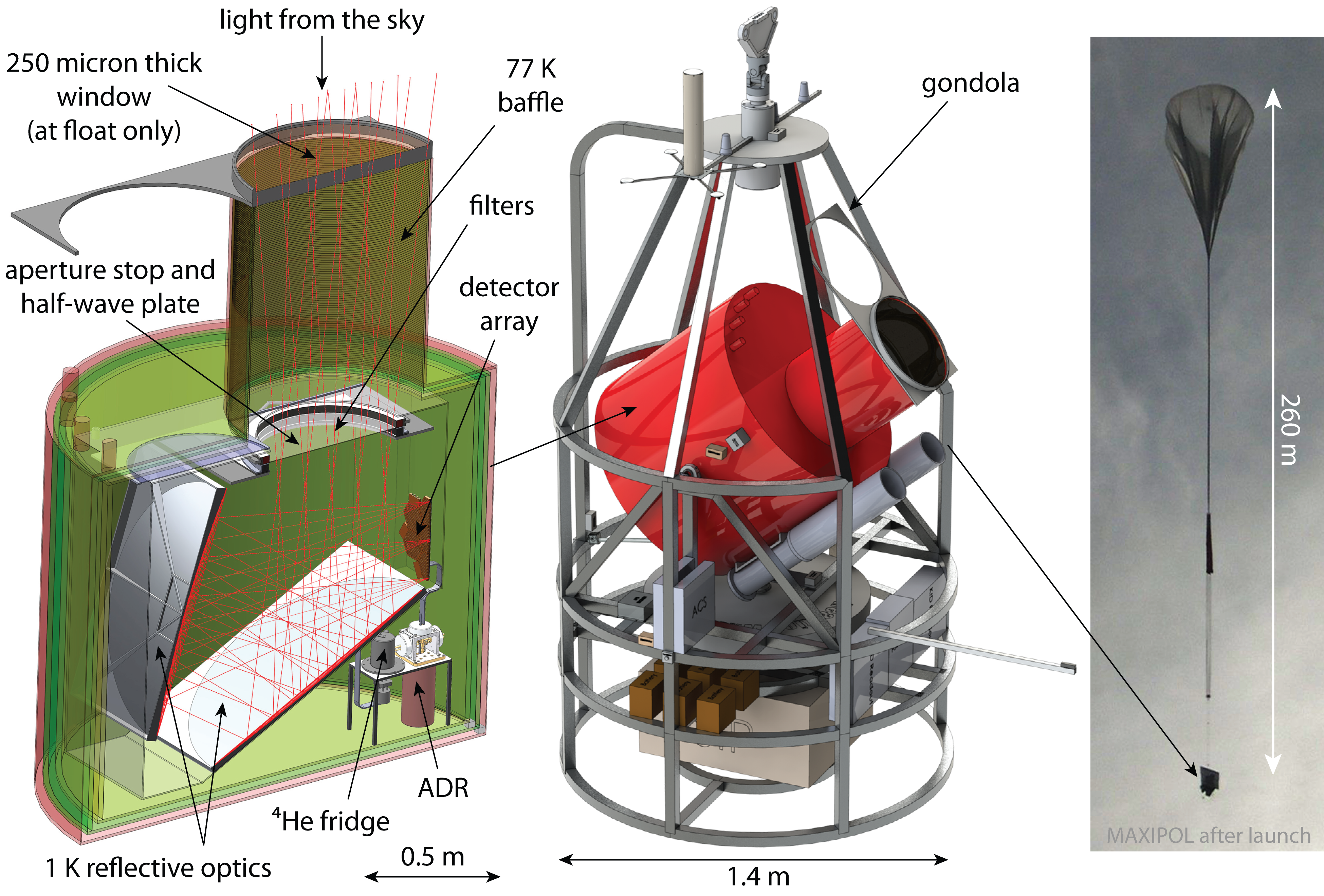}
\caption{\footnotesize{The balloon-borne \pname\ instrument.
    \textbf{Left:} A cross-sectional view of the polarimetric
    receiver.  \textbf{Center:} The receiver mounted in the gondola.
    \textbf{Right:} An example photograph of a balloon-borne
    instrument after launch.}}
\label{fig:instrument}
\end{figure}


\section{Instrument Details}
\label{sec:instrument}

The \pname\ polarimeter is based on a catoptric crossed-Dragone
telescope composed of an off-axis parabolic primary mirror and a
hyperbolic secondary mirror yielding an effective focal length of
1200~mm (see Figure~\ref{fig:instrument}).
A 500~mm diameter aperture stop, that is cooled to 1~K, produces a
15~arcmin FWHM beam at 150~GHz.
A metal-mesh half-wave plate (HWP) mounted at this aperture stop will
rotate at 7.4~Hz using a drive system based on a superconducting
magnetic bearing (SMB), which minimizes vibrations and eliminates
heating from stick-slip friction.
When paired with the downstream analyzer, which is part of the
detector array, the linearly polarized component of the sky signals
will be modulated at four times the rotation frequency of the HWP in
the data stream (29.7~Hz).
The detector array, which contains 2317 single-polarization,
horn-coupled, aluminum \lekids, is mounted at the focal plane of the
telescope.

To minimize detector loading and maximize sensitivity, the entire
polarimeter will be mounted inside a cryostat with liquid nitrogen and
liquid helium stages.
%
%
%
A box enclosing the optics will be thermally isolated from the liquid
helium stage of this cryostat by Vespel legs and maintained at 1~K
using a Chase Cryogenics closed-cycle $^4$He refrigerator, which is
capable of providing 500~$\mu$W of cooling power for five days.
The detector array is mounted inside this 1~K optics box using
thermally insulating Vespel legs, and it is cooled to 100~mK by a
two-stage adiabatic demagnetization refrigerator (ADR).
The ADR contains a ferric ammonium alum (FAA) paramagnetic salt pill
with a 0.12~J capacity when regulated at 100~mK, backed by a gallium
gadolinium garnet (GGG) salt pill with a 1.2~J capacity capable of
cooling to 800~mK.
Optical loading is controlled by (i) a radiation shield in the
``snout'' of the receiver held at 77~K, which is designed to control
far side-lobes, (ii) a second radiation shield held at the liquid
helium bath temperature ($\sim1.5$~K, see below) designed to terminate
power from the horns down to -20~dB, (iii) a series of low-pass
metal-mesh filters that reject high-frequency radiative power from the
sky, and (iv) the receiver uses a double window mechanism based on
designs that have flown successfully in the
past\cite{staggs:1996,ebex:2012} in which a retractable thick window
supports the differential pressure between the inside and outside of
the cryostat on the ground, while a thin (250~$\mu$m) window is
revealed at float.
%
%
The polarimeter operates in two configurations, one sensitive to a
single spectral band centered on 150~GHz and the other sensitive to
260 and 350~GHz bands.
The only differences between configurations are (i) the band-defining
filters, (ii) the single-moded waveguide section of the horn, (iii)
the aperture stop diameter, and (iv) the half-wave plate.

The polarimeter is housed in a robust gondola designed for multiple
flights.
The balloon and the instrument will reach an altitude of 100-125~kft,
where the ambient pressure is 5-13~mbar.
We regulate the nitrogen vapor to atmospheric pressure and allow the
helium bath to reach ambient pressure at float, resulting in a
nitrogen stage temperature of 77~K and a helium stage temperature of
approximately 1.5~K.
%


\begin{figure}[t]
\centering
\includegraphics[width=\textwidth]{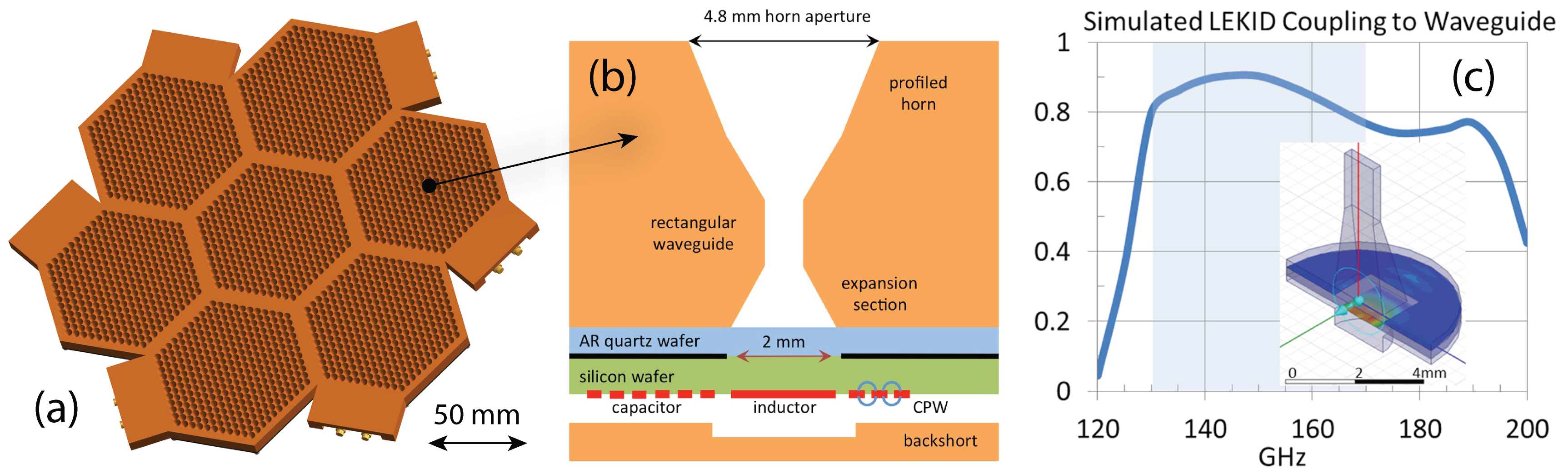}
\caption{\footnotesize{The \pname\ detector array. \textbf{Left:} The
    focal plane is composed of seven modules each containing 331
    horn-coupled detectors.  \textbf{Center:} A cross-sectional view
    of one focal plane pixel shows the profiled horn, the single-moded
    waveguide and the \lekid. \textbf{Right:}. A plot showing the
    detector absorptance as simulated by the Ansoft Software package
    HFSS.}}
\label{fig:detector_array}
\end{figure}


\section{\lekid\ Detector System}
\label{sec:detectors}

The \pname\ focal plane architecture is shown in
Figure~\ref{fig:detector_array}.
The focal plane diameter is $\sim300$~mm, and it consists of seven
hexagonal modules.
Each module contains 331 horn-coupled \lekids, for a total of 2317
detectors.
All of the detectors in a single module are frequency-multiplexed in a
135--270~MHz readout band, allowing the entire focal plane to be read
out using only seven SiGe bipolar cryogenic low noise amplifiers
(LNAs)\cite{bardin:2008} and seven pairs of coaxial cables.

We use horns to couple sky signals to the detectors.
As shown in Figure~\ref{fig:detector_array}b, the horn narrows down to
a single-mode rectangular waveguide section, which reduces spillover
and stray-light loading inside the optics box and provides
high-performance polarization selection and an integrated high-pass
filter.
The $1 F \lambda = 4.8$~mm horn aperture allows a dense focal plane
layout and yields high receiver sensitivity and fast mapping speeds.
%
With this relatively small horn aperture, the instrument beam is
primarily defined by the cold aperture stop.
The horns couple with approximately 38\% efficiency to the sky at
150~GHz, and the remaining 62\% of the horn beam is terminated on the
1~K optics box which results in negligible loading.
The final section of the waveguide is re-expanded to flatten the
propagation impedance at the low-frequency edge of the band, which
improves optical coupling and allows the radiation to be launched
efficiently into the silicon \lekid\ wafer.

\begin{figure}[t]
\centering
\includegraphics[width=0.9\textwidth]{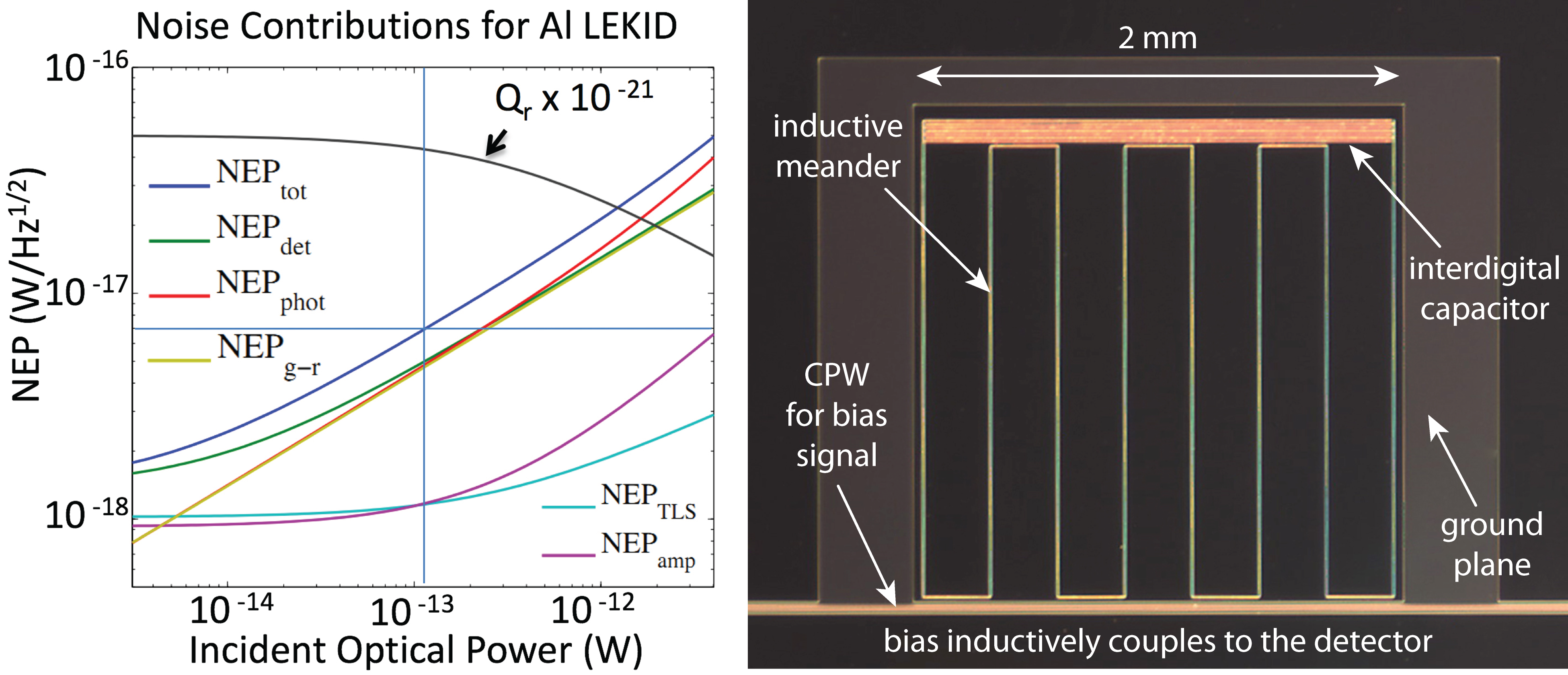}
\caption{\footnotesize{\textbf{Left:} Expected NEP and resonator $Q$
    versus incident optical loading. These curves were computed using
    both theory\cite{zmuidzinas:2012} and
    measurements\cite{mckenney:2013}.  The total NEP (NEP$_{\rm tot}$)
    curve includes contributions from the photon loading (NEP$_{\rm
      phot}$) and the expected detector noise (NEP$_{\rm det}$), which
    includes generation-recombination (g-r) noise, two-level system
    (TLS) noise, and readout amplifier (amp) noise.  The light blue
    vertical and horizontal rules indicate the expected loading and
    NEP for \pname.  The resonator $Q$ curve has been multiplied by a
    factor $10^{-21}$ for display purposes. \textbf{Right:} A
    photomicrograph of one \lekid.}}
\label{fig:detectors}
\end{figure}

The \lekids\ will be fabricated from 20~nm thick aluminum films
deposited on 300~$\mu$m thick, high-resistivity silicon substrates.
The baseline pixel design consists of a back-illuminated
single-polarization \lekid\ where the inductor/absorber is a meandered
aluminum trace on a silicon substrate with a filling factor designed
to match the wave impedance (see Figure~\ref{fig:detectors}).
This design is similar to the absorbers used in the
NIKA\cite{monfardini:2011} and MAKO\cite{swenson:2012} instruments.
An anti-reflection coating (ARC) of fused quartz is used to impedance
match the silicon wafer and the horn.
The ARC and detector wafers are mounted directly to the back of the
horn plate and a metal back plate closes each module.
The back plate has metal cavities behind each detector that are
designed to act as backshorts close to $\lambda/4$ in length.
%
%
The small fraction of the radiation that propagates laterally in the
two dielectric substrates will be absorbed by a titanium nitride (TiN)
film deposited and patterned on the ARC.
This TiN layer has apertures in it to allow radiation from the sky to
propagate from the horns through the quartz and silicon to the
detectors.
%


\begin{figure}[t]
\centering
\includegraphics[height=0.195\textheight]{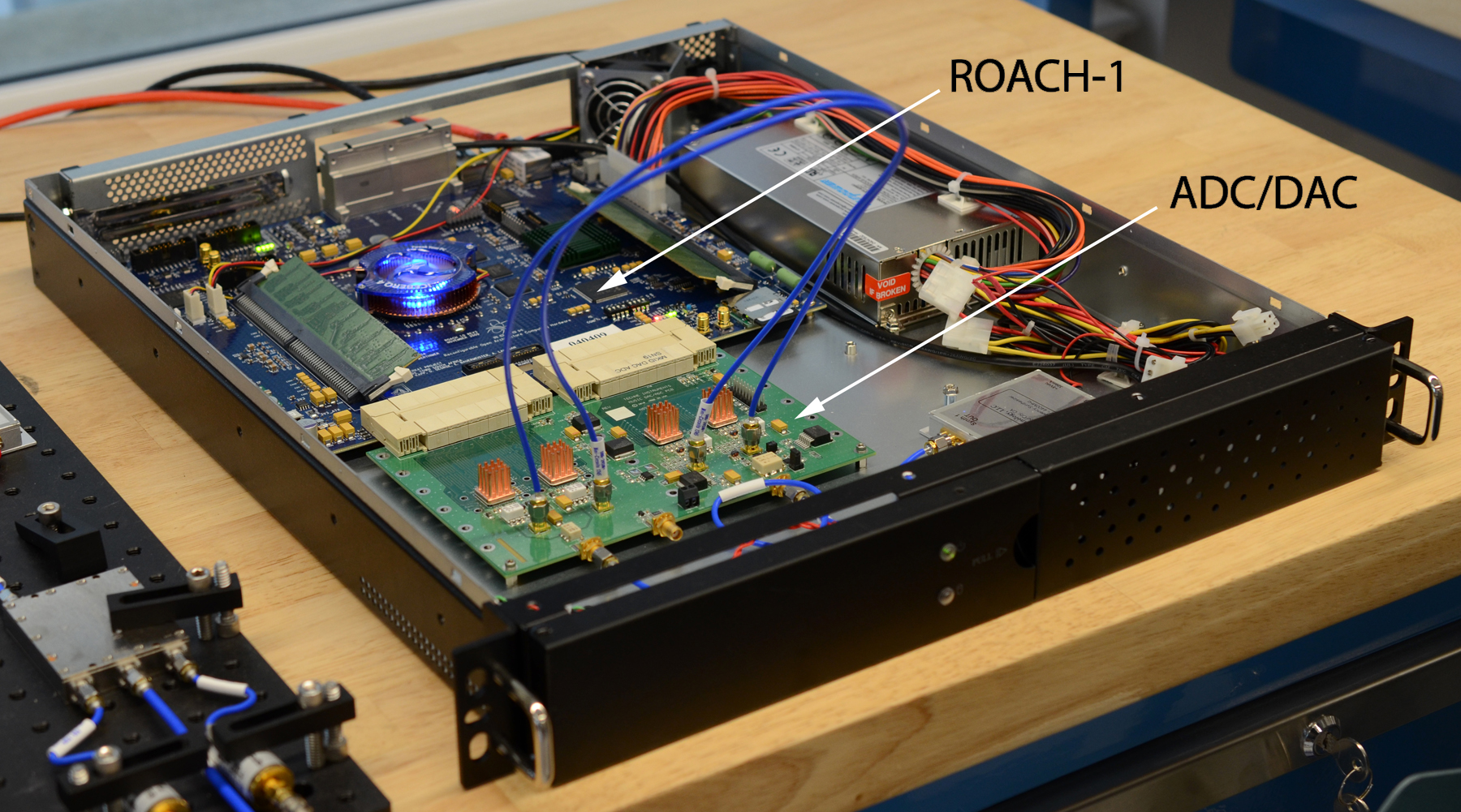}
\includegraphics[height=0.195\textheight]{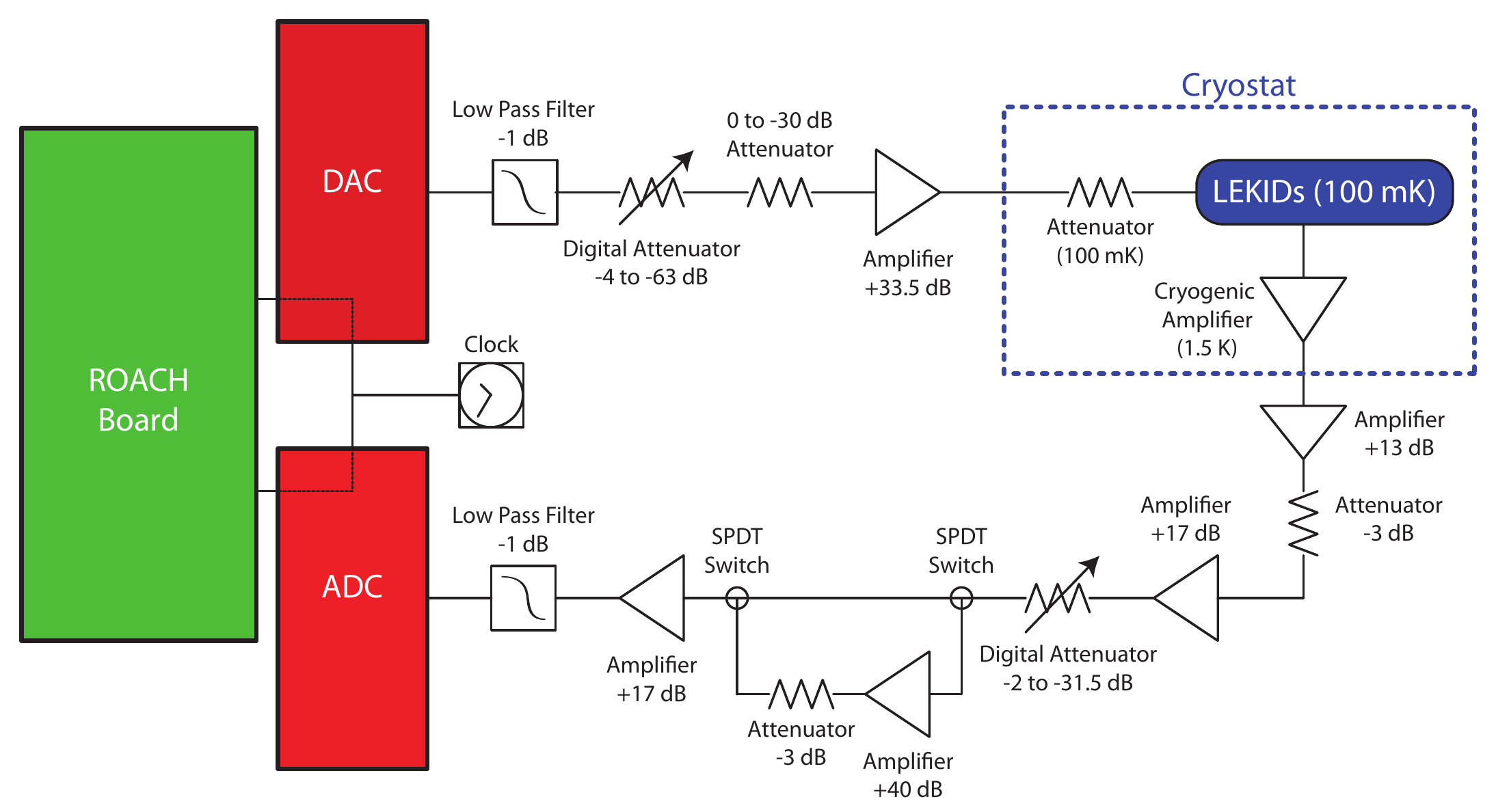}
\caption{\footnotesize{\textbf{Left:} A photograph of a ROACH-I
    readout currently running at Columbia. \textbf{Right:} A schematic
    of the \pname\ readout circuit.}}
\label{fig:readout}
\end{figure}

The detector readout electronics are based on (i) the ROACH signal
processing board, developed by the CASPER collaboration\cite{casper},
which hosts a Xilinx field-programmable gate array (FPGA) and (ii) the
successful Open-Source Readout system developed at Caltech for the
MUSIC instrument\cite{duan:2010}.
The stimulus tones for the \lekid\ resonators are directly synthesized
in the 135–-270~MHz band using a continuous playback ring-buffer which
drives a 16-bit digital-to-analog converter (DAC).
The output of the DAC is amplified and bandpass-filtered to suppress
harmonics before being fed to the detector wafer.
Digitally controlled attenuators are used to optimally adjust the
overall signal levels.
After exiting the cryostat, the modulated carrier tones are then
further amplified and filtered before being sampled by a 12-bit
analog-to-digital converter (ADC).
The digitized data is then channelized by the same FPGA.
The voltage data from 331 channels which contain resonators will then
be sent via Ethernet to data acquisition computers for storage.
%
%
%
%
Figure~\ref{fig:readout} shows an overview of the \pname\ data
acquisition electronics.


\section{Observation Plans}
\label{sec:observation}

One of the lowest foreground regions at 150~GHz anywhere on the sky is
located at Galactic coordinates (l, b) = (99$^{\circ}$, 75$^{\circ}$),
which is in the Northern sky\cite{finkbeiner:1999}.
To observe this region, \pname\ will be launched from the ESRANGE
balloon facility in Kiruna, Sweden (67.9$^{\circ}$~N,
20.2$^{\circ}$~E).
A balloon flight from Kiruna allows five days of integration time, and
this low-foreground target field can be continuously observed for the
entire mission.
%
%
%
%
%
%
The total time spent on the patch during a five-day flight is 103
hours, equivalent to an 86\% observing efficiency, and the total sky
coverage is 1133~deg$^2$.
The observation program consists of at least two, five-day flights.
The first flight will use the 150~GHz configuration, while the
second flight will use the 260 and 350~GHz configuration.
Instrument recovery and relaunch can be done quickly with Kiruna
flights, so two flights in two successive years are possible.
%


\begin{figure}[t]
\centering
\includegraphics[width=0.6\textwidth]{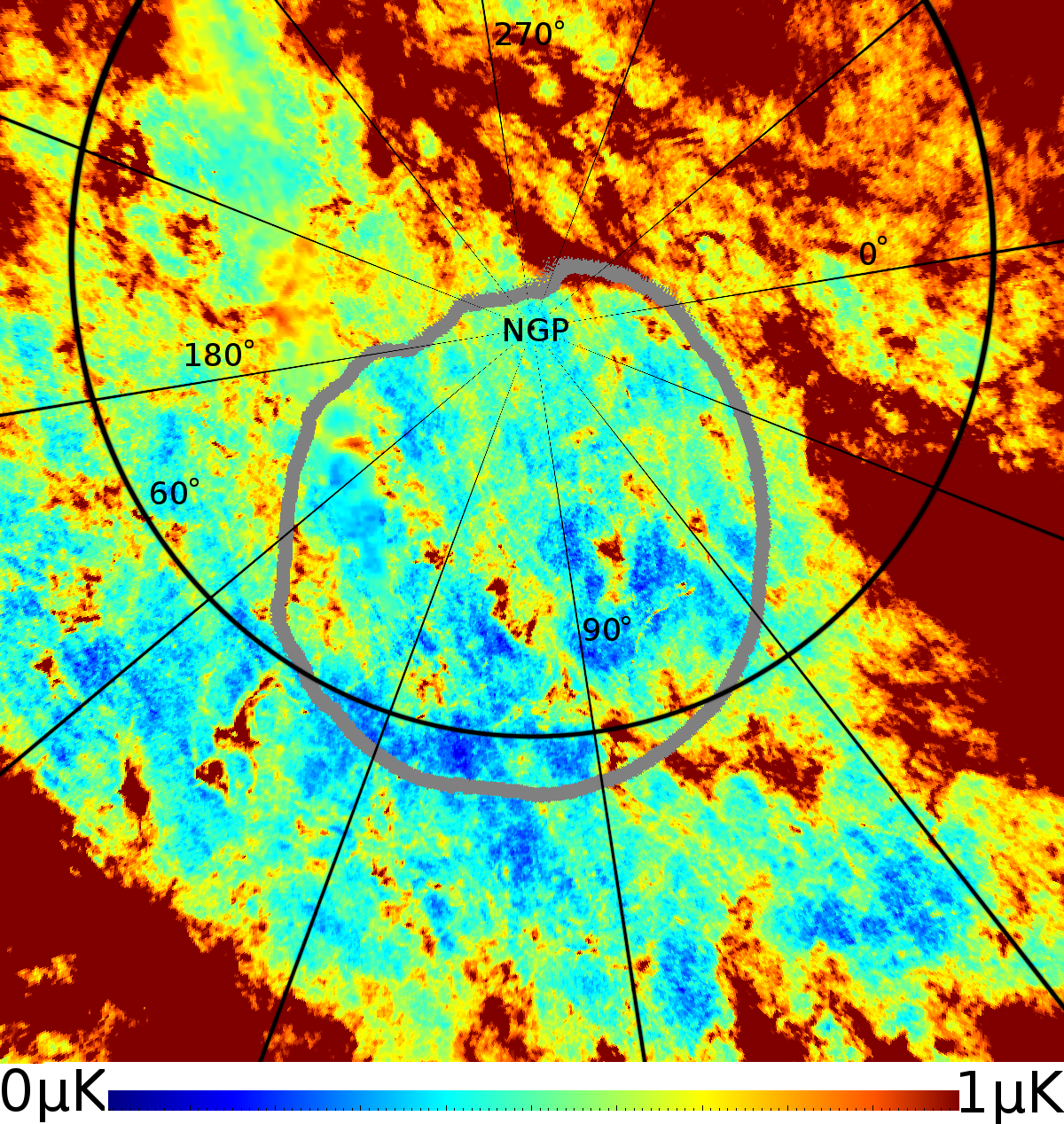} \\
\vspace{\baselineskip}
\includegraphics[width=0.8\textwidth]{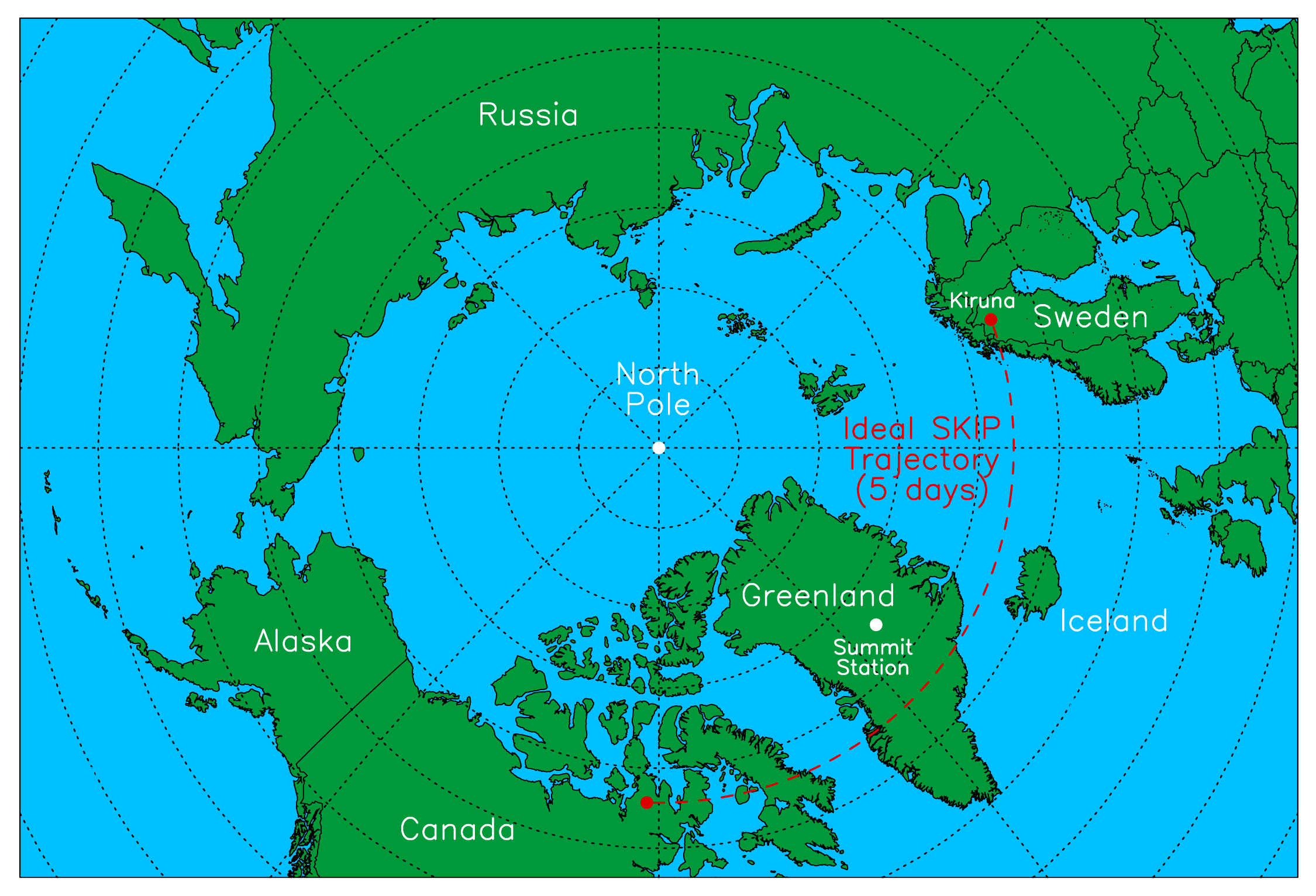}
\caption{\footnotesize{\textbf{Top:} Predicted polarized dust
    amplitude at 150 GHz in the Northern celestial sky, assuming a
    10\% dust polarization fraction~\cite{finkbeiner:1999}. The
    1133~deg$^2$ \pname\ target field is indicated by a gray
    boundary. \textbf{Bottom:} A map showing the anticipated
    \pname\ trajectory.  The balloon is launched in Kiruna, Sweden,
    and the instrument lands in Northern Canada five days later.}}
\label{fig:observing}
\end{figure}





\begin{thebibliography}{99}

\bibitem{baumann:2009} D. Baumann, et al., \textit{AIP Conference
  Series}, \textbf{1141}, 10, (2009).

\bibitem{eriksen:2008} H. K. Eriksen, J. B. Jewell, C. Dickinson,
  A. J. Banday, K. M. Gorski, and C. R. Lawrence,
  \textit{ApJ}, \textbf{676}, 10, (2008).

\bibitem{planck_cib:2013} The Planck Collaboration. {\it A\&A
  submitted}, (2013). arXiv:1309.0382

\bibitem{finkbeiner:1999} D. P. Finkbeiner, M. Davis, and
  D. J. Schlegel, \textit{ApJ}, \textbf{524}, 867, (1999).

\bibitem{staggs:1996} S. T. Staggs, N. C. Jarosik, S. S. Meyer, and
  D. T. Wilkinson, \textit{ApJ}, L1, (1996).

\bibitem{ebex:2012} B. Reichborn-Kjennerud, et al., \textit{SPIE
  Conference Series}, \textbf{7741}, (2010).


\bibitem{bardin:2008} J. C. Bardin and S. Weinreb, \textit{IEEE MWCL},
  \textbf{19}, 6, 407, (2008).


\bibitem{monfardini:2011} A. Monfardini, et al., \textit{ApJS},
  \textbf{194}, 24, (2011).

\bibitem{swenson:2012} L.~J.~Swenson, et al., \textit{SPIE Conference
  Series}, \textbf{8452}, (2012).


\bibitem{zmuidzinas:2012} J.~Zmuidzinas,
  \textit{Annu. Rev. Condens. Matter Phys.}, \textbf{3}, 169, (2012).

\bibitem{mckenney:2013} C.~McKenney, \textit{JOLT submitted}, (2013).

\bibitem{duan:2010} R. Duan, et al., \textit{SPIE Conference Series},
  \textbf{7741}, (2010).

\bibitem{casper} http://casper.berkeley.edu


\end{thebibliography}
\end{document}